\documentstyle[11pt,aaspp]{article}

\received{December 27, 1994}
\accepted{July 17, 1995}
\lefthead{Lilly et al.}
\righthead{Evolution of luminosity function to z $\sim$ 1}

\tighten

\begin{document}

\title{The Canada-France Redshift Survey VI:\\
Evolution of the galaxy luminosity function to z $\sim$ 1}

\author{S. J. Lilly\altaffilmark{1}}
\affil{Department of Astronomy, University of Toronto, Toronto, Canada }

\author{L. Tresse\altaffilmark{1} and F. Hammer\altaffilmark{1}} 
\affil{DAEC, Observatoire de Paris-Meudon, 92195 Meudon, France} 

\author{David Crampton\altaffilmark{1}} 
\affil{Dominion Astrophysical Observatory, National Research Council of Canada, 
       Victoria, Canada} 
 
\author{O. Le F\`evre\altaffilmark{1} }
\affil{DAEC, Observatoire de Paris-Meudon, 92195 Meudon, France}

\altaffiltext{1} {Visiting Astronomer, Canada-France-Hawaii Telescope, which
is operated by the National Research Council of Canada,
the Centre National de la Recherche Scientifique of France and the University of Hawaii}

\begin{abstract}

The cosmic evolution of the field galaxy population has been studied out to a redshift 
of $z \sim 1$ using a sample of 730 $I$-band selected galaxies, of which 591 have 
secure redshifts with median $<z> \sim 0.56$. The tri-variate luminosity function $\phi (M,color,z)$ 
shows unambiguously that the population evolves and that this evolution is strongly 
differential with color and, less strongly, with luminosity. The luminosity function of red 
galaxies shows very little change in either number density or luminosity over the entire 
redshift range $0 < z < 1$.  In contrast, the luminosity function of blue galaxies shows 
substantial evolution at redshifts $z > 0.5$. By $0.5 < z < 0.75$ the blue luminosity function 
appears to have uniformly brightened by approximately 1 magnitude. At higher redshifts, 
the evolution appears to saturate at the brightest magnitudes but continues at fainter 
levels leading to a steepening of the luminosity function. A significant excess
of galaxies relative to 
the Loveday et al. (1992) local luminosity function is seen at low redshifts $z < 0.2$ around 
$M_{AB}(B) \sim -18$ and these galaxies may possibly represent the descendants of the evolving blue 
population seen at higher redshifts. The changes seen in the luminosity function are also 
apparent in color-magnitude diagrams constructed at different epochs and in the $V/V_{max}$ 
statistic computed as a function of spectral type. Finally, it is argued that the picture of galaxy 
evolution presented here is consistent with the very much smaller samples of field 
galaxies that have been selected in other wavebands, and with the results of studies of 
galaxies selected on the basis of Mg II 2799 absorption 

\end{abstract}


\section{Introduction}

Deep redshift surveys, such as the Canada-France Redshift Survey (CFRS) described in 
earlier papers in this series (Lilly et al. 1995a CFRS I, Le Fevre et al. 1995 CFRS II, 
Lilly et al. 1995b, CFRS III, Hammer et al. 1995 CFRS IV and Crampton et al. 1995 
CFRS V), allow the direct study of galactic evolution with cosmic epoch. However, since 
different galaxies are observed at different epochs along the past light null-cone, the 
study of galaxy evolution is necessarily statistical in nature and requires a combination of 
knowledge about the population of galaxies at different epochs and an understanding of 
the physical processes occurring in individual galaxies which together produce the 
changes seen in the population. 

One of the most basic descriptions of the galaxy population is the luminosity function. 
This paper constructs the tri-variate luminosity function of galaxies, $\phi (M,color,z)$, over 
the $0.0 < z < 1.3$ redshift interval from the CFRS sample described in CFRS I-V.  We stress at 
the outset that this paper is concerned with describing the population of galaxies, defined 
at different epochs by various criteria, and no effort is made to determine the physical 
processes occurring in individual galaxies. Similarly, no assumptions are made, or 
conclusions drawn, as to whether individual galaxies move in or out of the various 
populations at different redshifts.  We seek simply to provide a description of the 
luminosity function at each epoch.

$H_0$ is taken to be 50 kms$^{-1}$Mpc$^{-1}$ 
throughout the paper and, except where noted to the 
contrary, we generally assume $q_0$=0.5.  Except where discussing the results of others, we 
have utilized the AB magnitude normalization scheme (Oke 1972), where $B_{AB} = B - 
0.17$, $V_{AB} = V$, and $I_{AB} = I + 0.48$.

\subsection{Previous work}

Until recently there has been very limited redshift data for field galaxies at 
cosmologically interesting redshifts.  Early studies of the nature of the galaxy population 
at significant look-back times (e.g. Kron 1980, Koo 1986) were based on comparisons of 
color and magnitude distributions with the predictions of models, integrated over redshift 
space. With many free parameters available in the models, it was always possible to find 
acceptable fits to the data and these models suffered from obvious problems of 
uniqueness. 

During the late 1980's, redshift data on limited samples of galaxies became available. In a 
landmark paper, Broadhurst et al. (1988) published a sample of 230 galaxies selected in 
the $B$-band ($B < 21.5$) which had a median $<z> \sim 0.2$.  These data were initially 
analyzed as before in terms of comparing the $N(z)$ distribution with modeled predictions. 
This showed the initially surprising result that the $N(z)$ distribution appeared to be the 
same as that predicted for ``no-evolution'' models of the galaxy population despite the 
apparent excess in the $N(B)$ number counts relative to these same models. This result was 
extended to $21.5 < B < 22.5$ by Colless et al. (1990) (149 galaxies, $<z> \sim 0.3$) and to $22.5 
< B < 24$, by Cowie et al. (1991) (12 objects with $<z> \sim 0.4$ - see also LCG). Subsequently, 
Eales (1993) (see also Lonsdale and Chokshi 1993) constructed luminosity functions from 
the $B < 22.5$ samples and confirmed that they were consistent with a uniform increase in 
the apparent number density of galaxies. 

If the faint end of the local galaxy luminosity function is flat, $\alpha \sim 1$, 
then an increase in 
the apparent number density requires, especially for higher $q_0$, either a dramatic 
luminosity evolution of some part of the galaxy population, a large degree of merging 
over relatively recent epochs or a change in the cosmological geometry. All of these 
processes were invoked to varying degrees by various authors (e.g. Lilly, Cowie and Gardner, 
1991, hereafter LCG, Babul and Rees 1992, Cowie et al. 1991, Yoshii and Fukugita 1992, 
Broadhurst et al. 1992, see for example Lilly 1993 and Koo and Kron 1992 for recent 
reviews). 

One significant problem with these relatively shallow samples is that the redshift baseline 
is sufficiently small that evolutionary effects can only be seen relative to the ``local 
population'', placing a premium on uniformity between the deep and local samples. 
Unfortunately, there are significant differences between the selection criteria for local 
determinations of the luminosity function and those based on deep surveys and, as 
emphasized by Ferguson and McGaugh (1995), these can lead to differences in the 
derived luminosity function.  Our new CFRS sample allows us to look for evolution 
within a single sample in which the selection criteria are essentially constant 
over a wide range of redshift.

Along with the $I$-band selected CFRS, smaller samples of faint 
galaxies selected at comparable depths in other wavebands are 
now becoming available. Glazebrook et al. (1995) have presented 84 objects with $22.5 < 
B < 24$ (analyzed to produce a luminosity function by Colless 1995) and Songaila et al. 
(1994) have a large sample of over 300 K-band selected galaxies spanning a very wide 
range of magnitudes which have been analyzed by Cowie et al. (1995).   However, in 
terms of galaxies at high redshifts, both of these samples are still quite small, each 
containing around 40-50 galaxies at $z > 0.5$ as against the 350 at $z > 0.5$ in the CFRS 
sample that is the subject of the present analysis (see Section 1.3 below). We return to 
compare these different samples in Section 4 below.

\subsection{The importance of $I$-band selection}

As noted above, much of the initial work in this area has been based on $B$-selected 
samples of galaxies (Broadhurst et al. 1988, Colless et al. 1990, LCG, Cowie et al. 1991, 
Colless et al. 1993), with the goal of understanding the steep $B$-band galaxy counts, and has 
been generally confined to redshifts $z < 0.5$. 

A basic difficulty encountered in B-selected samples at $z > 0.3$ is that the observed B-
band is redshifted progressively further into the ultraviolet where different types of 
galaxies have quite different properties. This is shown in Figure 1(a) which shows the color 
difference between observed-B and rest-B for three galaxy spectral energy distributions 
from Coleman et al. (1980, CWW). As can be seen in Figure 1(a), the $B$-band $k$-corrections 
for different galaxy spectral types span 4 magnitudes at $z \sim 0.75$. The large variation 
between different galaxy types has a number of undesirable consequences. First, 
predictions of what is expected, even in the simplest ``no-evolution'' scenario, depend 
sensitively on the assumed mix of galaxies as a function of absolute magnitude and the 
poorly constrained ultraviolet properties of each type of galaxy.  This has led to a 
lingering controversy as to the amount of evolution required by these $B$-selected data (see 
e.g. the discussions in Broadhurst et al. 1992, Lilly 1993, Koo and Kron 1992, Koo et al. 
1993).  
As an extreme example, Koo et al. (1993) were able to match the counts and 
redshift distributions of the faint $B$-selected samples with a completely unevolving 
galaxy population by altering the properties of the local population (in essence assuming 
its properties are unknown) and assuming the most favorable cosmological geometry.

Similar difficulties are encountered in the more direct approach of computing luminosity 
functions. Clearly, the $k$-corrections for individual galaxies must be determined with 
considerable precision in $B$-selected samples.  As described by Colless (1995), this can 
be reliably done from the observed $(B-R)$ colors, at least up to $z \sim 0.7$.  A more subtle 
problem, however, is that the wide range of $k$-corrections within the population, or even 
within any reasonable sub-division by color, means that at a given observed $B$ magnitude 
different classes of objects will populate the rest $B$-band luminosity function at very 
different luminosities.  Put another way, the faintest bins in absolute $B$ magnitude 
(presumably produced at any redshift by the faintest galaxies at the survey limit) will 
contain only the bluest objects since the redder objects will have,  at this same apparent $B$ 
magnitude, a much brighter absolute $B$ magnitude. The large range in $k$-corrections acts 
to effectively eliminate red objects from most of the luminosity function.

Selection in the $I$-band substantially reduces the problems of working at high redshift 
($0.3 < z < 1.0$).  At $z \sim 0.5$ and at $z \sim 0.9$, the observed $I$-band corresponds to the rest-frame 
$V$-band and $B$-band respectively. Thus the selection of the distant sample is as well 
matched as possible to that of most local samples of galaxies. Correspondingly, the 
$k$-corrections to produce a $B$-band luminosity function are minimal and do not depend 
strongly on spectral type (and not at all at $z = 0.89$) . This is illustrated in Figure 1(b) which 
shows the color between observed-$I$ and rest-$B$ for the three spectral energy distributions 
plotted in Figure 1(a).  Following from this, the mix of galaxy types at a given $M_{AB}(B)$ is not 
biased in terms of color.  

\subsection{The CFRS sample}

As an unprecedentedly large and deep $I$-band selected redshift survey, the CFRS is 
particularly well-suited to the determination of the luminosity function at high redshift. 
In addition to the $I$-band selection, many aspects of the CFRS program were specifically 
designed with this goal in mind.  

The statistically complete CFRS sample consists of 943 objects in five fields selected 
without regard to color or morphology or environment.  Of these 943 objects, 
730 are not stars 
or quasars (defining these spectroscopically rather than morphologically) and of these 
galaxies, redshifts have been secured for 591 (81\%) with a median $<z> = 0.56$.  The 
reliability of these identifications has been confirmed by a large number of repeat 
observations (see CFRS III) and the nature of the remaining unidentified galaxies can be 
inferred in most cases, at least in a statistical sense, from their photometric properties and 
from analysis of the identifications of a subset of objects that were initially unidentified 
but for which repeated observations secured a redshift (see CFRS V and section 2.4). 

The surface brightness selection effects in the original photometric sample are well 
understood and should be minimal for most types of galaxy (CFRS I). The identifications 
in the spectroscopic sample have been shown to be unbiased in surface brightness relative 
to the original photometric sample (CFRS IV). 

An additional attractive feature of the CFRS sample is that the sample spans five 
magnitudes $17.5 < I_{AB} < 22.5$ in the same areas of sky, and the objects that were 
observed spectroscopically were selected from this range without regard to apparent 
magnitude (CFRS II). Thus, at any redshift, the bright and faint ends of the luminosity 
function are determined from exactly the same volumes of space.  

\section{Calculation of the luminosity function}

We have computed luminosity functions in the rest-frame $B$-band from the statistically 
complete sample of 730 galaxies defined in CFRS II-V using a simple $1/V_{max}$ formalism 
(Felten 1976). The luminosity functions were calculated and cross-checked
using two independently written
codes (by SJL and LT).

\subsection{The $k$-corrections}

The $(V-I)_{AB}$ color straddles the rest-frame $B$-band 
for $0.2 < z < 0.9$. Thus, for each galaxy, we 
first assign a spectral type by comparing the observed $(V-I)_{AB}$ colors, available for all 
objects, with those computed at that redshift from the spectral energy distributions given 
by CWW.  As in Lilly (1993) an interpolation scheme is used whereby the CWW 
elliptical spectral energy distribution is assigned to spectral type 0, their Sbc to type 3, 
their Scd to type 5 and their Irr to type 6 with other spectral types representing 
interpolations between these. This spectral type (a real number) then defines the complete 
spectral energy distribution by interpolation from the CWW spectral energy distributions. 

Since we wish to compute an absolute magnitude in the $B$-band from $I$-band 
observations, we then compute a ``$k$-correction color'' for each galaxy from its assigned 
rest-frame spectral energy distribution. This $k$-correction color is computed from the 
interpolated spectral energy distribution as $(B-I_z)_{AB}$, where the subscript $z$ indicates that 
the waveband is shifted to shorter wavelengths by a factor of (1+$z$).  The $k$-correction 
color is in effect a modeled interpolation of the observed $(V-I)_{AB}$ color for the redshift 
range $0.2 < z < 0.9$ and is the quantity plotted 
in Figure 1(b) for three spectral energy distributions.  

The absolute magnitude in the rest-frame $B$-band is then computed based on the isophotal 
$I_{AB}$ magnitude, the distance modulus calculated from the luminosity distance, the 
bandwidth stretching term and the $k$-correction color $(B-I_z)_{AB}$:

\[M_{AB}(B) = I_{AB} - 5 \log (D_L/10 pc) + 2.5 \log (1+z) + (B-I_z)_{AB}\]

It should be noted that the last two terms could be re-written as a conventional k-
correction for the $I$-band plus a rest-frame $(B-I)_{AB}$ color derived from the spectral energy 
distribution.  The luminosity distance is defined as:

\[D_L = (c/H_0) (1+z) Z_q(z)\]

where $Z_q(z)$ is the function

\[Z_q(z) =  \frac{(q_0 z + (q_0 -1)((1+2q_0 z)^{0.5} -1 )))}{q_0^2 (1+z)}\]

As noted above, since the $(V-I)_{AB}$ color straddles the rest-frame $B$ for $0.2 < z < 0.9$, this 
procedure should lead to very small uncertainties in the computation of $M_{AB}(B)$ over this 
range, and none at all at $z \sim 0.2$ and at $z \sim 0.9$ (when the observed-$V$ or observed-$I$ 
corresponds to rest-$B$).  For some purposes it has proved convenient to characterize the 
assigned rest-frame spectral energy distribution by the rest-frame $(U-V)_{AB}$ color. This 
roughly corresponds to the observed $(V-I)_{AB}$ at $z \sim 0.5$. 

\subsection{Calculation of the luminosity function}

Since the form of the luminosity function at high redshift was unknown, the luminosity 
function was computed directly in bins of absolute magnitude, color and redshift space 
using the $1/V_{max}$ formalism.

\[\phi (M,color,z) dM = \sum_{k} \frac{1}{V_{max}}\]  

The sum is carried out over all galaxies in the sample lying within the specified range of 
luminosity, color and redshift. The $V_{max}$ is computed as the comoving volume within 
which each galaxy (as defined by its absolute magnitude and assigned spectral energy 
distribution) would remain in the sample in question, i.e. would satisfy the limits in 
apparent magnitude of the survey $17.5 < I_{AB} < 22.5$ and/or the limits in redshift of the 
bin. Writing the upper and lower limits of the redshift bin as $z_U$ and $z_L$ and 
the redshifts at which the object in question would have $I_{AB}$ = 17.5 and $I_{AB}$ = 22.5 as 
$z_{17.5}$ and $z_{22.5}$ respectively, then we have:

\[V_{max} = (c/H_0)^3 \int_{\max(z_{L},z_{17.5})}^{\min(z_{U},z_{22.5})
} \frac{Z_{q}^{2}(z)}
{(1 + z)(1+2 q_0 z )^{1/2}}\]

The volume is based on an effective solid angle, $d \Omega$, of the survey of 112 arcmin$^2$ 
(CFRS II)

\subsection{Binning in absolute magnitude}
 
A magnitude limited sample such as the CFRS has widely varying numbers of objects 
with different luminosities. In order to avoid having very small numbers of objects in 
some bins or unnecessarily losing luminosity resolution where there were many objects, 
we adopted a scheme whereby the size of the magnitude bin was allowed to vary so as to 
include roughly equal numbers of objects in each bin.

As noted above, the range of $k$-correction colors at a given redshift within the galaxy 
population (even if it is sub-divided into color-bins) means that, unless the observed 
waveband exactly matches the desired rest-frame waveband (so that the $k$-correction 
color is zero for all spectral types) the highest and lowest bins in absolute magnitude will 
be populated by only a subset of the galaxy population (i.e. the reddest or bluest objects 
within the population or color-bin).  In an $I$-band selected sample, this effect is modest, 
and is zero at $z$ = 0.89 if the luminosity function is calculated in the rest-frame $B$-band.  
To avoid any such problems we have simply eliminated the faintest few tenths of a 
magnitude in setting the bins for the luminosity functions at higher and lower redshifts. 
As shown in Figure 1(b) the range of $k$-correction increases roughly linearly as the redshift 
changes away from $z$ = 0.89, so the amount of the luminosity function eliminated in this way 
was simply set to $0.6 \times |(z-0.89)|$. 

\subsection{Calculation of uncertainties}

Uncertainties in the luminosity functions were calculated using a bootstrap algorithm in 
which the galaxy sample was resampled in a series of 1000 Monte Carlo simulations. In 
each simulation, a new sample of galaxies equal in number to the original sample was 
generated by randomly sampling the galaxies in the original sample (i.e. allowing 
both
duplications and omissions of some of the original galaxies).  This method takes into 
account the variable weighting of objects in the $1/V_{max}$ algorithm but does not take into 
account any effects of the clustering of galaxies in redshift space.  

Assuming the clustering is independent of absolute magnitude the small scale clustering 
of galaxies in redshift space leads to an effective reduction in the number of independent 
galaxies and this will lead to an increased uncertainty (by a factor of approximately  
$\sqrt{2.5}$ 
- see CFRS VIII) in the density normalization $\phi *$ of any given redshift bin.  In practical 
terms, the statistical uncertainty in $\phi *$ arising from the numbers of objects, $N^{-0.5}$, 
is much 
smaller than the uncertainty that arises because of correlated uncertainties with the other 
parameters in any fitting of the luminosity function.

\subsection{Treatment of unidentified sources}

The CFRS complete sample of 943 sources contains 200 stars and 6 quasars (which were 
ignored for the present analysis), 591 confirmed galaxies with redshifts and 146 
unidentified objects. Seven of the unidentified objects are very likely to be stars based on 
their morphologies and colors (see CFRS V) and these were also eliminated from further 
consideration. The unknown nature of the remaining 139 galaxies represents a basic 
uncertainty in constructing the luminosity function. These objects are found towards the 
faint limit of the survey. The identification rate for galaxies (i.e. 
once the known stars and quasars are removed) as a function of $I_{AB}$ magnitude
is shown in Figure 7 of CFRS V.  

The uncertainty introduced by the unidentified objects has been addressed in three 
different ways.  We have first simply ignored the unidentified galaxies. This results in a 
``minimal'' luminosity function that is based on only those galaxies that were actually 
identified. This is clearly the most conservative treatment, but results in a luminosity 
function that must at some level be an underestimate of the true luminosity function.  
Secondly, we have followed the usual procedure of weighting the galaxies with known 
redshifts by the inverse of the fractional identification rate of the sample (calculated after 
the stars and quasars were removed), determining this as a function of magnitude. 
In essence, this scheme incorporates them into the sample making the implicit 
assumption that they have the same color and redshift distribution as the identified 
galaxies in the sample at the same magnitude. We refer to this as the ``weighted'' luminosity 
function. 

This weighting scheme seems unnecessarily restrictive because, as discussed in detail in 
CFRS V, we have a good idea of what the unidentified objects are, from two separate 
lines of argument. First, in the course of the project, 99 unidentified objects were re-
observed and identifications subsequently secured for approximately 70\% of these 
objects (see CFRS III). If we make the assumption that these 99 re-observed objects are 
similar to the roughly 110 objects which were unidentified and which were not re-
observed, then we can say that the redshift distribution of about half of the final list of 
139 unidentified objects is known, at least statistically, and is given by the redshift 
distribution of the ``recovered failures''.  Second, as described in CFRS V, we devised a 
simple redshift estimator based on an empirical comparison of the $(V-I)_{AB}$ and $(I-K)_{AB}$ 
colors, the $I_{AB}$ magnitude and the compactness of the images, with a learning sample 
given by galaxies with secure redshifts. This estimator successfully reproduced the 
redshift distribution of the ``recovered failures''. It was then applied to all 139 
unidentified objects in the final catalogue, with the one minor, and ad hoc, alteration that 
20 galaxies were arbitrarily placed at $z > 1$ where our observational set-up created a 
strong bias against absorption line objects (see CFRS V for a detailed discussion).  Our 
third approach to calculating the luminosity function was thus based on placing the 
unidentified objects at their estimated redshifts. This obviously means that we cannot 
consider the luminosity function for these red galaxies at $z > 1$.  We refer to this third 
estimate as the ``best estimate'' luminosity function.  

\subsection{Representation of the luminosity function}

As with other flux-density limited samples, most of the galaxies in the CFRS
occupy a small range of apparent magnitude and thus the typical absolute magnitudes in
the sample are correlated
with distance. At very low redshifts, this effect is dealt
with when constructing local luminosity functions 
through a straightforward volume correction and the resulting
luminosity function is defined over a wide range of luminosities.  In the case of a
deep sample such as the CFRS,
the range of distances sampled corresponds to a wide range of cosmic epochs, and thus,
at each epoch, only a limited range of luminosities is present in the sample.  A simple
volume correction is innappropriate if evolution in the population is a possibility.
 
Consequently only
a relatively short segment of the luminosity function can be determined at each epoch
from the CFRS.  
This makes it dangerous to fit the usual Schechter function - the individual parameters 
will be poorly determined and the uncertainties in the parameters are
inevitably highly coupled. Furthermore, the parameters of the Schechter function 
(such as the ``faint end slope'', $\alpha$, or the ``knee'', M*) that 
result when the function is fitted 
over a small range of luminosities (as would be the case here) respond to features that may be quite 
unconnected with their intuitive meanings and thus the changes of the parameters with 
redshift could be quite misleading.

The philosophy adopted in the present work is to compare the luminosity functions
at different epochs as non-parametrically as possible (i.e. directly point by point)
rather than to fit analytic functions
and interpret changes in the parameters of these fits. However, as an aid in
translating our results from diagram to diagram and to enable other workers
to reproduce our luminosity function on their own diagrams, we have fitted 
analytic functions to our luminosity functions.  We could have chosen any functional
representation, such as a polynomial, but chose to use the Schechter function.
The fits were simple $\chi ^2$ minimizations on the binned luminosity function, 
with the model being integrated across each luminosity bin.  

The parameters of these various fits (without formal uncertainties) 
are listed in Table 1, along with the range 
of $M_{AB}(B)$ over which 
they are valid. {\it We stress that these are representations of our data over a limited 
luminosity range, and are not intended to be 
``determinations'' of the Schechter parameters}. The  
changes in individual parameters, such
as $\alpha$, $M*$ and $\phi *$, should not be viewed in isolation.

\section{The population of galaxies $0.0 < z < 1.3$}

\subsection{The trivariate luminosity function}

The CFRS sample is sufficiently large that we can subdivide the sample into several
redshift bins and into two color bins and still have reasonable numbers of objects
with which to construct the tri-variate luminosity function $\phi (M,color,z)$.

We show in Figures 2(a)-(c) four computations of the luminosity function where the sample 
has been split into blue and red populations (by rest-frame color - 
dividing the sample at the spectral energy 
distribution of the CWW Sbc galaxy, roughly the median spectral type) and into 
five redshift intervals: $0.05 < z < 0.2$, $0.2 < z < 0.5$, $0.5 < z < 0.75$, $0.75 <
z < 1.0$
and $1.0 < z < 1.3$ (blue galaxies only). 
The choice of 
these redshift bins is to some degree arbitrary but was motivated as follows:  In an $\Omega = 1$ 
cosmology, the first two bins correspond to intervals in cosmic time of 0.2 $\tau_0$ centered on 
0.85 and 0.65 $\tau_0$ respectively. The next two redshift bins correspond to intervals of about 
0.1 $\tau_0$ centered on approximately 0.5 and 0.4 $\tau_0$. It should be noted that the upper 
magnitude limit of $I_{AB} > 17.5$ excludes galaxies with $M_{AB}(B)$ brighter than between $-20.5$
and $-21.0$ (depending on their color) from the 
lowest redshift $0.05 < z < 0.2$ bin. The excluded region 
of the luminosity function is to the left of the vertical lines in the top two panels in each Figure
2(a-c). Figure 2(a) shows 
both the ``minimal'' luminosity function and the ``weighted'' luminosity function for $q_0$ = 0.5, 
Figure 2(b) shows the ``best estimate'' luminosity function for $q_0$ = 0.5 and Figure 2(c) shows the 
``best estimate'' luminosity function for $q_0$ = 0.   

In each of the panels at $z > 0.2$ in Figure 2, we plot the segment of the Schechter function 
that has been fit to the luminosity function in that redshift range as a solid curve. 
We also
show in each panel, the fits to the $0.2 < z < 0.5$ blue and red 
luminosity functions (dashed curves) along with the Loveday et al. 
(1992) local luminosity function (dotted curve). 
It should be noted that this latter is derived from Loveday et 
al's whole galaxy population, not split by morphological or spectral type and is provided 
primarily as a reference point within the diagram.  With the important caveats 
discussed in Section 2.6 (i.e. that these are {\it representations} of the data not
{\it determinations} of the parameters) 
the parameters of the Schechter segments are given in Table 1.

Comparison of Figures 2(a)-(c) shows that the qualitative features of the luminosity function 
are largely independent of the treatment of the unidentified objects (and to a certain 
degree, of the value of the deceleration parameter $q_0$).   In detail, it can be seen 
that the ``best estimate'' luminosity function is slightly steeper than the ``minimal'' one since 
the unidentified objects are concentrated at the faint end of the sample and thus at the 
faint end of the luminosity function at any redshift. An additional small effect 
is that the increase relative to the ``minimal'' luminosity 
function is larger at $z > 0.5$, reflecting the fact that the distribution of estimated redshifts 
is shifted slightly to higher redshifts than that of the sample as a whole (CFRS V).  

In what follows we concentrate on Figure 2(b) which represents our best estimate of the 
luminosity function for $q_0$ = 0.5, before seeing how the situation would change if $q_0 \sim 0$.  
There are three interesting results apparent in Figure 2(b) which we describe 
phenomenologically in the next three sub-sections. We then present two further analyses 
that illustrate these effects in Sections 3.2 and 3.3 before discussing, in Section 3.4, 
the possible origin and cosmological context of these results.

\subsubsection{The local luminosity function: is there evolution at $0.0 < z < 0.2$?}

The CFRS was designed to produce a homogeneous sample of galaxies at $z > 0.2$ and
there are only 50 galaxies at $z < 0.2$. However,
there is clear evidence in Figure 2, for a population of faint galaxies ($M_{AB}(B) \sim 
-18$) in the lowest redshift bin $0.05 < z < 0.2$ (which we 
might expect to be a ``local'' population), 
that has a significantly higher comoving number density than in the local luminosity 
function of Loveday et al. (1992), an effect also noticed in their smaller sample by Tresse et al. 
(1993). Comparison of the upper two levels on Figure 2(a-c) shows that this low redshift, 
low luminosity, population is well matched by our luminosity functions at $0.2 < z < 0.5$.
Indeed, {\it within our sample} we see no evidence {\it per se} for evolutionary changes
in the
galaxy population between the 
$0.05 < z < 0.2$ and $0.2 < z < 0.5$ redshift bins, although there are significant differences with the
Loveday et al. (1992) local luminosity function.

As noted above, our luminosity function at $z < 0.2$ contains no luminous galaxies 
with $M_{AB}(B)
< -20.5$ because of the bright end limit of our sample ($I_{AB} > 17.5$), and so we cannot state 
categorically whether the differences with Loveday et al. (1992) 
represents a difference in the overall normalization of the 
luminosity function (i.e. a different $\phi$*) or a difference only at the faint end although we 
strongly suspect the latter.  

This excess at faint magnitudes is similar in size, though a little higher
in luminosity, to that which appears 
in the luminosity function derived from the CfA sample by Marzke et al. (1994a). The 
excess evident in their Figure 9(b) at $M \sim -15.75$ would appear in our figure at $M_{AB}(B) 
\sim -17.5$ 
once allowance is made for the different values of $H_0$ used and the zero-point offset 
between $B$ and $B_{AB}$.  In both samples an excess of approximately a factor of four over 
the Loveday et al. (1992) luminosity function is observed.

There are very few published local luminosity functions that are split by color. 
The most directly comparable is that quoted by Metcalfe et al. (1991).
We show in Figure 3 our $0.05 < z < 0.2$ and $0.2 < z < 0.5$ luminosity functions, 
split as in Figure 2 by the implied rest-frame $(U-V)$ color (though at these low redshifts
our $(V-I)$ observed color more closely matches the rest-frame $(B-R)$) and the
local luminosity function, split at $(B-V) = 0.72$, derived by Metcalfe et al. (1991). 
These are extremely similar.
The steep slope of the luminosity function
of bluer galaxies is also
seen indirectly in Marzke et al.'s (1994b) decomposition of the CfA luminosity function
into different morphological bins.

A fair conclusion at this point
is that uncertainties in the local luminosity function are sufficiently
large as to make the determination of whether there is evolution back to $z \sim 0.2$ 
(c.f.
Broadhurst et al. 1988) rather hard. At higher redshifts, we have the advantage that the
luminosity function can be constructed from our single CFRS sample removing many of these
ambiguities and so evolutionary changes are much more secure.  

\subsubsection{No change in the luminosity function of redder galaxies $0.2 < z < 1.0$}

The luminosity function of the redder galaxies (redder than the present-day CWW Sbc 
spectral energy distribution) represented in the panels on the left of Figure 2(b)
clearly shows 
remarkably little change with epoch back to the highest redshifts encountered, $z \sim 1$. The 
fit to the $0.2 < z < 0.5$ luminosity function is clearly a reasonable representation of the 
luminosity function of red galaxies at $0.5 < z < 0.75$ and $0.75 < z < 1.0$ and,
as noted above, within the accessible 
range of luminosities, also at $0.05 < z < 0.2$.  This presumably implies that the population of 
ellipticals and bulge dominated spirals is roughly constant with epoch. While one can 
imagine combinations of evolutionary processes that could produce the appearance of a 
static population as in Figure 2(b), there is little evidence {\it per se} for either a substantial 
decrease with increasing redshift in the numbers of these redder galaxies, as might be 
expected if the population had been formed recently through the merger of massive 
actively star-forming sub-units, or for a brightening of more than a few tenths of  
magnitude from the passive evolution of the dominant old stellar populations in these 
galaxies (c.f. Tinsley and Gunn 1976, Bruzual 1983, Yoshii and Takehara 1988, Bruzual 
and Charlot 1993).  We return to this below.

In order to try to quantify this lack of change in the luminosity function of the redder 
galaxies, a Schechter function was first fit to the luminosity function of these galaxies 
over the whole range $0 < z < 1$.  This yielded $\alpha = -0.5$, $M_{AB}*(B)$ = -21.00 and 
$\phi *$ = 
0.00185 Mpc$^{-3}$.  Keeping $\alpha$ fixed at this value, but allowing M* and 
$\phi *$ to vary, Schechter functions
were then fit to the red luminosity functions in the three redshift bins $0.2 < z 
< 0.5$, $0.5 < z < 0.75$ and $0.75 < z < 1.0$.  The 1$\sigma$ error ellipses for the two free 
parameters in these fits are shown in Figure 4.  It should be noted that the fit for $z > 0.75$ is 
least well constrained because nearly all the galaxies have $M_{AB}*(B) < -21.0$ at these 
redshifts. 

Clearly, these fits are consistent with no change at all in either $M*$ or $\phi *$ over the whole 
redshift range. However, they are also consistent with a modest brightening with look-back
time if accompanied by a decrease in the total number of objects. Over the interval 
from $z \sim 0.3$ to $z \sim 0.6$, a brightening of at most 0.5 magnitudes and a decrease of at most 
33\% in the comoving space density of these red galaxies would be acceptable.

\subsubsection{Evolution in the luminosity function of bluer galaxies $0.2 < z < 1.0$}

In contrast to the luminosity function of red galaxies, the luminosity function of the bluer 
galaxies (i.e. bluer than the present-day CWW Sbc) shown in the right hand panels of Figure 
2(b) shows significant changes as the redshift increases. In general terms, this may be 
characterized by a brightening of the luminosity with look-back time, although the same 
effect could, of course, also be produced by an increase in comoving density.  In terms of 
luminosity evolution, this appears to be differential with luminosity and to ``saturate'' at 
the bright end, producing a steepening of the luminosity function. Between $0.2 < z < 0.5$ 
and $0.5 < z < 0.75$, the luminosity function in Figure 2(b) brightens by about 1 magnitude, 
while in the next step to $0.75 < z < 1.0$ there is no change in the bright end but a further 
brightening of about 1 magnitude around $M_{AB}(B) \sim -20$.  The saturation at bright 
magnitudes is evident also in the final $1.0 < z < 1.3$ bin, which shows little change 
relative to the $0.75 < z < 1.0$ bin for $M_{AB}(B) < -21.5$.

Because our redshift baseline is so large, the evolution of the galaxy luminosity function 
is evident within our single sample and is thus independent of any uncertainties in the 
definition of the ``local'' population. We regard Figure 2 as providing incontrovertible 
evidence that the galaxy population changes with cosmic epoch and believe that the ``no-
evolution'' scenario advocated by Koo et al. (1993) is thus no longer tenable.

\subsubsection{Results with $q_0$ = 0}

The effect of changing $q_0$ to lower values is straightforward.  There is little change at the 
lower redshifts, and at higher redshifts, the different populations have lower comoving 
densities but higher luminosities, i.e. they tend to move diagonally down and to the left 
in the diagrams.  

For the red galaxies, this increases the amount of passive luminosity evolution that is 
permitted since the high redshift population will be intrinsically more luminous.  It also 
allows a larger increase in the comoving density since $z \sim 1$, because the comoving 
density of the red population will be lower for $q_0 \sim 0$ at the earlier epochs.   We have 
represented these changes as vectors emanating from the centers of the error ellipses in 
Figure 4.  

For the blue galaxies, the effect of moving the luminosity functions down and to the left 
has little effect on the amount of evolution required. There is still a requirement for the 
equivalent of approximately one magnitude of luminosity evolution to $z \sim 0.6$, and for 
additional evolution to $z \sim 0.9$ at $M_{AB}(B) \sim -21$.

\subsection{The rest-frame color-magnitude diagrams at different epochs}

In Figure 5, color magnitude diagrams are plotted for galaxies in the sample in different bins 
of redshift.  The absolute $B$ magnitude calculated as above is plotted against the 
rest-frame $(U-V)_{AB}$ color computed for the interpolated spectral energy distribution that 
matches the observed $(V-I)_{AB}$ color. As noted above, these colors are exactly equivalent 
at $z \sim 0.5$ and the transformation at other redshifts is modest. The unidentified galaxies 
with estimated redshifts (see above and CFRS V) are shown as open symbols.

In a magnitude-limited sample such as the CFRS, the mean absolute magnitude increases 
with redshift due to the bright and faint sample limits and to the obvious volume effects.  
If attention is confined to objects brighter than $M_{AB}(B)$ = -20 (one magnitude below 
present-day L*), then the increase with redshift of the luminosity of the blue population 
of galaxies, relative to the red population, seen on Figure 3, is clearly apparent in Figure 5.
The well-defined color-luminosity relation present at low redshifts is eliminated at the
higher redshifts by the emergence of the population of bright blue galaxies.

\subsection{$V/V_{max}$ as a function of spectral type}

As has been known for many years (Schmidt 1968) the statistics of the locations of 
objects within their observable volumes $V_{max}$, computed as above, gives information on 
the evolution of that population.  Values of 0.5 indicate a homogeneous unevolving 
population while values greater than 0.5 indicate a population whose comoving density, 
at a given luminosity, is increasing with distance. In order to see which spectral types are 
producing the evolution seen in Figure 2 and Figure 5, the $V/V_{max}$ statistic has been computed 
for galaxies in the redshift range $0.3 < z < 1.0$ as a function of spectral type, with the 
results shown in Table 2 for $q_0$ = 0.5 (the $V/V_{max}$ statistic is insensitive to $q_0$). This 
redshift range was chosen since it is the range over which the spectral typing of galaxies 
from their $(V-I)_{AB}$ colors is most accurate and over which there should be little bias in 
the determination of redshifts as a function of spectral type (see CFRS IV and CFRS V 
for a discussion).  
 
The values of $V/V_{max}$ show a clear increase towards bluer galaxy spectral types, 
increasing from about 0.5 for galaxies redder than our spectral type 3 (i.e. the Sbc galaxy 
of CWW) to greater than 0.6 for the bluest galaxies. This is consistent with the changes 
seen in the luminosity function (Figure 2b) and color-magnitude diagrams (Figure 5).

\subsection{Discussion: the evolving galaxy population}

The above analyses show that the galaxy population in the $0.2 < z < 1$ interval is 
characterized by strongly differential evolution.  The luminosity function of red galaxies 
changes little if at all, while the luminosity function of blue galaxies brightens 
considerably, coming up from low luminosities to eventually dominate the luminosity 
function around present-day L*. Furthermore, within the blue population, there is 
evidence that the evolution is differential in luminosity leading to a steepening of the 
luminosity function.  This picture broadly matches the heuristic model for the evolving 
galaxy population constructed by Lilly (1993) to match the observed counts in $B$, $I$ and $K$, 
and the then available $N(z)$ distributions in $B$ and $I$. 

The significance of the unevolving red population is that in general 
terms, it supports the idea that massive, quiescent galaxies have been around in the 
Universe for a considerable time and got most of their activity over with at early epochs, 
beyond the reach of present surveys. This is consistent with what is known about early 
type galaxies in clusters to $z \sim 1$ (see e.g. Dressler et al. 1985, Rakos and Schombert 
1995).

The detailed interpretation of the lack of change in L* of the red population
requires knowledge of the strength of the so-called
passive evolution that arises, even in populations formed in a single burst
of stars, due to the changing number of turn-off stars and red giants.
The strength of the passive evolution
is strongly dependent on the slope of the initial mass function
(Tinsley and Gunn 1978) and the behaviour 
will be different for systems that continue to form stars.  For instance, in the models of 
Yoshii and Takahara (1986) and Bruzual and Charlot (1993), while single burst models 
fade by approximately 0.3 mag at a wavelength of 2 $\micron m$ as they double in age, models 
with a constant star-formation rate increase in brightness by the same amount. Thus the 
luminosity evolution of composite stellar populations may be quite mild and, for 
example, stellar populations with an exponentially declining star-formation rate with e-
folding time of order 7 Gyr have almost no luminosity change at 2.2 $\micron m$.
Given the sensitivity to the poorly constrained initial mass function, we regard the
amount of passive evolution as a quantity to be determined observationally, say from
observations of the fundamental plane in clusters of galaxies at high redshift, if
at all possible.

The nature of the evolving blue population is more intriguing.  Although it would be 
straightforward to view this as arising from the brightening of individual objects by 
modest amounts (1-2 magnitudes) it should be stressed that the luminosity function tells 
us only about the evolution of the population (defined in some observable way) and not 
about the evolution of individual objects or about the movement of objects into or out of 
the population in question.   Our analysis can not distinguish in a rigorous way between 
luminosity evolution and density evolution.

However, in the context of this ambiguity, the ``excess'' of low luminosity galaxies above 
the flat ($\alpha \sim 1$) faint end of the Loveday et al. (1992) luminosity function, seen here and 
in the CfA luminosity function of Marzke et al. (1994) is highly significant.  These 
galaxies are attractive descendants for the bright blue galaxies seen with high number 
density in our luminosity function at high redshift, $z > 0.5$.  It is important to appreciate 
that many of the more exotic interpretations that were advanced to account for the large 
numbers of blue galaxies seen even at quite bright magnitudes ($B < 24$) (such as bursting 
mini-haloes, Babul and Rees 1992, or wholesale merging of major mass 
concentrations, Broadhurst et al. 1992)  were introduced to a large degree
because of the difficulty 
of identifying a local population of sufficient number density if the local population was 
described by an $\alpha \sim 1$ flat luminosity function with low $\phi *$. Thus, it is noteworthy on 
Figure 2b that, within our whole sample, the highest comoving number density (per 
magnitude) is encountered in the lowest redshift $0.05 < z < 0.20$ bin at $M_{AB}(B) \sim -18$.  
Although these low redshift, low luminosity, galaxies do not themselves contribute 
significantly to the total count at our magnitude level, they do represent a ready source 
population for the evolving brighter galaxies seen at $z > 0.5$ (with $-21 < M_{AB}(B) < -20$) 
which do dominate the sample.

Establishing whether the evolution in the population seen in Figures 2(b) and 5 is indeed due 
to a simple brightening with look-back time of the individual objects that are seen today 
as a population of moderate luminosity objects ($M_{AB}(B) \sim -18$) will require detailed 
examination of their properties, and particularly of their morphologies and internal 
kinematics. Our investigations along these lines will be presented in due course 
elsewhere (e.g. Schade et al., 1995, CFRS IX). 
At this point, we simply note that the relatively normal morphologies, spectra 
and colors of the blue galaxies at $z > 0.5$ do not at first sight support many of the more 
exotic models in which they represent a population that has disappeared by the present 
epoch. We also note that the clustering properties of these galaxies are consistent
with those of local populations of star-forming galaxies if there has been
plausible growth of large scale structure in the Universe (Le Fevre et al. 1996, CFRS VIII, Hudon
and Lilly 1996)

One final comment can be made about the practicality of the volumetric test for $q_0$ 
discussed by Loh and Spillar (1986). The CFRS sample is very similar in size and depth 
to the sample used by Loh and Spillar in which essentially all of the redshifts were 
estimated photometrically by comparing colors with model spectral energy distributions.  
The demonstration of strongly differential evolution of the luminosity function in color 
(and probably also in luminosity) in the present work will make any future application of 
this particular cosmological test difficult unless individual masses can be determined for 
a large number of galaxies.  If it is assumed that the red galaxies form a stable population 
of fixed comoving number density (we believe that such an assumption would be quite 
unwarranted at the present time) then the values of $\phi *$ derived in Section 3.1.1 for the red 
galaxies alone clearly favor a value of $q_0$ larger than 0 (see Figure 4). Making this 
assumption, it is found that $q_0 < 0.1$ is formally excluded with 90\% confidence.  We 
ascribe little weight to this result.

\section{Comparisons with other samples}

The CFRS is considerably larger than any other current sample of faint galaxies, and this 
has allowed us to define the luminosity function to greater redshifts and with 
considerably finer sampling in color and redshift space than other studies. We review 
here the recent results of others based on smaller samples selected in different ways and 
argue that a  broadly consistent picture emerges.

To facilitate comparisons with these other samples, 
the upper panel in Figure 6, shows the overall ``minimal'' and ``weighted'' 
luminosity functions constructed from the entire CFRS sample integrated over 
all $0.0 < z < 1.3$ and undifferentiated by color. A Schechter function fitted
to the minimal luminosity 
function in the upper 
panel of Figure 6
gives $\alpha$ = -0.9 +/- 0.1 and $M_{AB}*(B)$ = -21.2 +/- 0.15. These are remarkably close to the 
values found by Loveday et al. (1992) from the DARS local redshift survey ($\alpha$ = -0.97 
and $M_{AB}(B)$ = -21.0 respectively) however our value of $\phi *$ (0.0041 +/- 0.0004 Mpc$^{-3}$) is 
about 2.3 times as high as the Loveday value. The Loveday et al. (1992) Schechter 
function is shown as the dotted line in Figure 6. The lower panels in Figure 6 show the ``best 
estimate'' CFRS luminosity function 
in the four redshift 
bins, 
$0.05 < z < 0.20$, $0.20 < z < 0.50$, $0.50 < z < 0.75$, and $0.75 < z < 1.0$ (as in
Figure 2b but now not split in color).  

The most important point in Figure 6 is that the 
{\it differential} evolution of the blue luminosity
function seen in Figures 2 and 5 relative to the static red luminosity function, can 
produce effects on the {\it integrated} luminosity function which can be described as 
either a steepening of the faint end slope, $\alpha$, or an increase in 
normalization $\phi *$, if the color-split is not made.

\subsection{The K-band selected sample of Cowie et al. (1995)}

Cowie et al. (1995) have constructed a cumulative luminosity function from a set of K-
selected samples of about 367 galaxies, including about 50 galaxies at $z > 0.5$. In broad 
terms, K-selection and I-selection should be more or less equivalent over the redshift 
range of interest since there is a relatively narrow range of $(I-K)_{AB}$ colors within the 
galaxy population. For instance, there is a variation of only
about one magnitude between the $(I-K)_{AB}$ 
colors of the CWW E and Irr spectral energy distributions at $z = 0.8$.   Cowie et al.'s 
(1995) Figure 2 compares the cumulative luminosity function in a single $0.1 < z < 1$ bin 
with a local sample based extensively on the Mobasher et al. (1993) sample. This figure is 
thus most directly comparable with the upper panel in Figure 6 (although the median 
redshift of our sample is rather higher - 0.56 as against 0.3). Cowie et al. note that the 
dominant effect in their sample must be a rise in the effective density $\phi *$, 
since the 
asymptotic luminosity density increases by a factor of 2.5 (for $q_0$ = 0.5) while the L* 
remains almost constant. They conclude that individual galaxies were slightly 
fainter in the past (0.25 +/- 0.2 mag) and that the 
number density rises as $(1+z)^{2.7}$. They infer that this requires that galaxies must be 
either combining through merging or 
else ``disappearing'' in some way as cosmic time progresses. They favored the first 
hypothesis because of the passive evolution that would would have been expected to make galaxies 
brighter in the past, noting that the increase in luminosity of
the {\it population} as a whole matched these evolutionary expectations whilst the lack of increase in
{\it individual objects} (i.e. in L*) did not. The issues raised by passive evolution were
discussed above in Section 3.4.

From our own larger sample, however, it is clear that 
separating the population into red and blue 
components as on our Figure 2(b) shows that the apparent increase in $\phi *$ comes about through 
the apparent brightening of the luminosity function of blue galaxies.  This produces a 
large number of luminous blue galaxies with luminosities comparable to present-day $L*$. 
Although the change in the population need not of course be produced by a 
corresponding evolution in individual objects, we note that quite modest changes to 
individual objects (1 to 2 magnitudes of dimming over 0.5$\tau_0$) are sufficient to account for 
the change in the population, especially if there is the low luminosity population above 
the Loveday et al. (1992) luminosity function  The numerous blue galaxies seen at $z > 0.5 $
do not need to ``disappear'' and can in fact fade by quite modest amounts to reproduce the 
observed effects.  While we are thus unable to categorically rule out large scale merging 
of galaxies we do not believe that it is required.

\subsection{The $B$-band selected sample of Colless (1995)}

Colless (1995) has constructed a preliminary 
luminosity function from the various B-selected samples 
extending down to the $22.5 < b_J < 24$ sample of Glazebrook et al. (1995). It is best 
defined at $z < 0.5$ since there are only 40 objects at $z > 0.5$. As in Broadhurst et al.
(1988), the changes in this luminosity function have been described in terms of an 
increase in the faint end slope of the luminosity function.
Qualitatively it shows similar behaviour to the lower four panels of
Figure 6. Perhaps of most interest 
in the present context is that in the $B$-selected samples analyzed by Colless (1995), the 
``excess'' at low luminosities appears between $z \sim 0$ and $z \sim 0.2$ and is the cause of the 
original effect discussed by Broadhurst et al. (1988).

We address below (Section 5.2) the specific question of whether there is any 
observational inconsistency between the marked paucity of high redshift galaxies at $z > 
0.7$ in published B-samples and the large number (33\% of the sample) at these redshifts 
found in the $I$-band selected CFRS sample. 

\subsection{The Mg II absorption sample of Steidel et al. (1995)}

Although small (58 objects at $0.3 < z < 1$, 70\% with spectroscopically confirmed 
redshifts) the sample of Steidel et al. (1995) is particularly interesting because it is based 
on the galaxies producing Mg II 2799 absorption lines in background quasars and thus 
represents a quite different selection methodology.

Steidel et al. (1995) concluded that neither the mean absolute magnitude nor the mean 
rest-frame (B-K) color of the absorbing galaxies changes over this redshift interval $0.3 < 
z < 1.0$.  The constancy of the mean color is at first sight surprising given the changes in 
the CFRS population seen in Figures 3 and 5 above. It cannot be argued (c.f. Steidel et al. 
1995) that our evolving blue population simply has too low an absolute K magnitude to 
produce absorption lines, since all of the CFRS galaxies at $z > 0.5$ in fact have $M_{AB}(K) < 
-20$, i.e. they lie above the threshold for producing absorption lines in the Steidel et al. (1995) 
sample.

To study this question in more detail, we have plotted on Figure 7 the $B$-band absolute 
magnitudes (calculate for $q_0$ = 0.5) of the Steidel et al. (1985) absorbing sample as a 
function of redshift (using data kindly provided by Dr. Steidel). Following our analyses 
above, a linear regression of $M_{AB}(B)$ on $z$ was separately computed for (a) the 28 red
galaxies with rest $(B-K)_{AB} > 1.3$, (b) the 25 blue galaxies with rest $(B-K)_{AB} < 1.3$ and 
finally (c) for the subset of 9 bluest galaxies with $(B-K)_{AB} < 1.0$. These lines are shown 
on Figure 7. While the absolute magnitudes of the red galaxies are indeed constant, the best 
fit to the two blue samples indicate an increase in the average luminosity between $0.3 < z 
< 1.0$ of 1 magnitude for the blue sample and 2 magnitudes for the very bluest objects. 

Although these effects are of marginal significance in the small Steidel et al. (1995) 
sample, they are entirely consistent with the changes seen on our 
Figures 2 and 5 above. It is clear on Figure 7 that the color-magnitude relation, which is 
clearly apparent in the absorbing sample at $z < 0.5$, breaks down at higher redshifts, 
exactly as seen in our Figure 5. We thus believe that, at present, the two samples are entirely 
consistent.

\section{N(z) and count extrapolations to fainter magnitudes and for other wavebands}

The $V_{max}$ approach to calculating the luminosity can be easily modified to produce 
predictions for the counts and redshift distributions at fainter magnitudes and in different 
wavebands.   As described in LCG, for each galaxy observed in our own sample for 
which a $V_{max}$ was calculated as above, the apparent magnitude as a function of 
redshift (placing the galaxy at all redshifts $0 < z < \infty$) is calculated in any waveband of 
interest from the known absolute magnitude and rest-frame spectral energy distribution.  
The contribution to the counts and/or $N(z)$ distribution at this apparent magnitude is then 
simply $dV/dz$ divided by the $V_{max}$ obtained as above for the galaxy in our $I$-band sample. 
This analysis is thus 
based on the implicit assumption that the observed population in our $17.5 < I_{AB} < 22.5$ 
sample does not evolve at either higher or lower redshifts. For simplicity we use 
the ``best estimate'' sample for this exercise. 

The number counts in the $B$-band and $I$-band that are generated in this way from the 
CFRS sample are shown in Figure 8 compared with number counts from the literature (taken 
LCG and references therein 
- see Koo and Kron 1992 for a review). In the next section we discuss the $I$-band extrapolation
and then turn to the $B$-band extrapolation.

\subsection{Extrapolation in the $I$-band}

By definition, the CFRS-based no-evolution prediction must match well the $I$-band 
counts around $I_{AB} \sim 22$ and the small discrepancy at this magnitude arises purely from 
the fact that stars are excluded from the prediction but included in the deep counts. As 
discussed by LCG, the flatter slope at brighter magnitudes is simply a reflection of the 
evolution already observed in the sample (e.g Figure 2(a-c)) - the sample contains high redshift 
objects at $z > 0.5$ that are not present at low redshift. 
At fainter magnitudes, the extrapolated counts fall below the observed 
counts, implying that there is continuing evolution of some sort to fainter magnitudes. At 
$I_{AB} \sim 24$ (two magnitudes below the CFRS) the observed count is 1.6 times that 
predicted from replicating the observed CFRS population to higher redshifts without any 
evolution.  

The modest shortfall at fainter magnitudes suggests that the predicted $N(z)$ 
from this model may be quite useful in estimating the redshift distributions of very faint I-
band selected galaxies that lie beyond the reach of current spectrographs. Topical studies 
in this area include measurements of the small distortions of background galaxies by 
intervening mass concentrations (e.g. Kaiser and Squires 1993) and the interpretation of 
the morphologies of the faintest galaxies revealed on very deep Hubble Space Telescope 
images. We show in Figure 9 the $N(z)$ predicted from this analysis for the $17.5 < I_{AB} < 22.5$ 
range of the CFRS and two fainter one magnitude slices of $I_{AB}$.

\subsection{Extrapolation in the $B$-band}

In the $B$-band, the shortfall at fainter magnitudes is more severe (see Figure 8).
Indeed, even at $B \sim 24$, 
there is already a shortfall of around 30\%. This implies that a $B < 24$ sample 
contains a number of very blue galaxies that are not represented in the $I_{AB} < 22.5$
selected CFRS. This should not be surprising because, as shown in LCG, the 
median $(B-I)_{AB}$ colors fall sharply, by at least 0.5 magnitudes, between $I_{AB}$ = 22.5 and 
$I_{AB}$ = 24.5 implying a rapid increase in the relative numbers of very blue galaxies with 
increasing depth in the $B$-selected samples.  A slightly deeper $I$-band sample would 
contain many more of these very blue galaxies that would have $B \sim 24$. We return to this 
point below. At the faintest magnitudes there is a shortfall of a factor of three and any 
$N(z)$ generated from this analysis is of limited use.

At this stage, we can only speculate on the nature of the evolving population that fills in 
the steep number counts.  Given what we have seen in the evolution of the bright blue 
galaxies, it may well be that these very faint objects will turn out to be at moderate 
redshifts and to represent the faint end of the luminosity functions shown in Figure 2(b).
  
One of the more interesting aspects of the $N(z)$ distributions from B-selected samples has 
been the marked paucity of galaxies at $z > 0.7$. The combined $21 < b_J < 22.5$ sample of 
Colless et al. (1990, 1993) has 4\% of galaxies at $z > 0.7$ with an unidentified fraction of 
only 6\% while the deeper $22.5 < b_J < 24$ sample of Glazebrook et al. (1995) has 13\% with 
a much larger 30\% unidentified fraction.   In contrast the $I$-band selected CFRS sample 
has approximately 33\% of galaxies at $z > 0.7$.  The $22.5 < B < 24$ sample of Cowie et al. 
(1991) has only 12 objects, of which one has $z > 0.7$ (L. Cowie private communication) 
although there are two more with $0.65 < z < 0.7$.

The Colless et al. (1990, 1993) samples are sufficiently bright that we can directly 
construct an equivalent sample from the CFRS $17.5 < I_{AB} < 22.5$ sample (since virtually 
no objects have $(b_J - I_{AB}) < 0$) at least in the three fields for which we have $B$ 
photometry.  This $21 < b_J < 22.5$ sub-sample of the CFRS contains 95 galaxies of which 
we have secure identifications for all but three (i.e. a 97\% completeness).
Seven galaxies have $z > 0.7$ (7\%) 
indicating no significant observational inconsistency at this level.

In order to assess the situation at the fainter levels, we used the $V_{max}$ formalism 
described above to predict $N(z)$ distributions for a $22.5 < B < 24$ sample.  It should be 
stressed that the analysis presented here should account for the different visibility of 
different classes of objects in the different wavebands but does not incorporate any 
evolutionary effects with redshift.  With this caveat in mind, the fraction of galaxies 
predicted at $z > 0.7$ for $22.5 < B < 24$ is 30\%, considerably higher than seen hitherto in the $B$ 
samples at this depth.  However,
it should be noted that there is a large fraction of 
unidentified objects (~30\%) in the full Glazebrook et al. (1995) sample (a subset of 30 
objects has only 10\% unidentified).  The explanation probably lies in a combination of 
effects. 
The very 
blue galaxies that are present in a $B < 24$ sample but absent from the $I_{AB} < 22.5$
CFRS  may well be at relatively low redshifts and, if this is the case, this will 
decrease the fraction of high redshift galaxies. Furthermore, the fraction of galaxies 
expected at $z > 0.7$ is increasing rapidly with depth. As discussed in CFRS I, our faint 
isophotal level for photometry ensures that even for these faint objects we 
include nearly all of the light whereas other photometry schemes may miss the equivalent 
of a few tenths of a magnitude. Perhaps the most likely 
explanation is that the existing $B$-selected samples may be systematically 
missing the
high redshift objects, perhaps because of limited spectral range.

We are currently extending the CFRS survey by observing blue objects so as to allow the 
direct construction of deeper B-selected samples. The results of this will be reported 
elsewhere.

\section{Summary}

The cosmic evolution of the field galaxy population has been studied over the redshift 
interval $0 < z < 1.3$ using the 730 galaxies (median $<z> \sim 0.56$) in the $I$-band selected 
CFRS redshift survey. The evolution of the population is best defined in terms of the tri-
variate luminosity function $\phi (M,color,z)$. The sample is large enough and spans a wide 
enough range in redshift and look-back time that evolutionary effects can be seen within 
the sample, thereby eliminating reliance on knowledge of the local population.

Three results have been found that indicate strongly differential evolution within the 
galaxy population.

(a)	The luminosity function of red galaxies (defined as redder than a typical present-
day Sbc galaxy) shows very little change in either number density or luminosity 
over the entire redshift range $0 < z < 1$, an interval corresponding to the last 
2/3 of the 
age of the Universe (for $\Omega \sim 1$). The luminosity function of the red galaxies is 
consistent with no change at all with $z$.  Between $z \sim 0.8$ and $z \sim 0.3$, a change in 
luminosity of at most a few tenths of a magnitude and a change in density of at 
most 33\% is indicated.

(b)	The luminosity function of blue galaxies (i.e. bluer than the present-day Sbc) 
shows substantial evolution at redshifts $z > 0.5$. At $0.2 < z < 0.5$, these galaxies 
are concentrated at low luminosities.  By $0.5 < z < 0.75$ the population appears to 
have uniformly brightened by approximately 1 magnitude (although this could 
also be described as an increase in comoving density). At higher redshifts, the 
evolution appears to saturate at the brightest magnitudes but continues at fainter 
levels leading to a steepening of the luminosity function. At the highest redshifts 
$1.0 < z < 1.3$, we sample only the bright end of the luminosity function and no 
additional evolution is seen.

(c)	At low redshifts $z < 0.2$, a significant excess relative to the Loveday et al. (1992) 
local luminosity function is seen around $M_{AB}(B) \sim -18$, similar (though slightly 
brighter) to that found by Marzke et al. (1995) in the local CfA survey. Our
color-dependent local luminsoity function is very similar to that of Metcalfe et al (1991).
This numerous
population, which may have evolved since $z \sim 0$, may represent the
descendants for the evolving blue population seen at higher redshifts
after modest luminsoity evolution. This population
would remove some of the motivations for introducing an exotic 
population to account for the numerous blue population seen at $z > 0.5$.

The changes seen in the luminosity function are also apparent in color-magnitude 
diagrams constructed at different epochs
and in the $V/V_{max}$ statistic computed as a function 
of spectral type. Given that the 
evolution is seen within a single sample, we regard this as secure evidence for changes in 
the galaxy population and believe that the no-evolution hypothesis (e.g. Koo et al. 1993) 
is no longer tenable.  

It is argued that the picture of galaxy evolution presented here presents no inconsistencies 
with the very much smaller samples of field galaxies that have been selected in other 
wavebands ($B$ and $K$), or with the results of studies of galaxies selected on the basis of 
Mg II 2799 absorption

\acknowledgements

	The CFRS project would not have been possible without the support of the 
directors of CFHT and of our two national TACs (CTAC and CFGT).  We have 
benefitted during the course of this project from conversations with many colleagues and 
particularly Ray Carlberg, Richard Ellis, Chuck Steidel, Mike Fall and Len Cowie.  
We are also grateful to the referee for his careful reading of the manuscript.
Chuck Steidel kindly provided his data to us in electronic form.
SJL's 
research is supported by the NSERC of Canada and the project has been facilitated by a 
travel grant from NATO.

\clearpage

\begin{figure}

\caption{Values of the $k$-correction color (see text) required to produce absolute 
magnitudes in the rest $B$-band from observations in the $I$-band (left panel) and $B$-band 
(right panel) as a function of redshift for three different spectral types of galaxies. Curves 
are for the CWW E spectral energy distribution (top), the CWW Sbc (middle) and CWW 
Irr (bottom).}

\end{figure}

\begin{figure}

\caption{(a) The ``minimal'' and ``weighted'' luminosity functions (see text) for the CFRS 
sample split by redshift (vertically as indicated by the label in each panel) and intrinsic 
color (redder than CWW Sbc on the left, bluer on the right). Redshift range and number 
of objects in that range are indicated by the label in the upper left of each panel. 
The vertical bar in the $0.05 < z < 0.2$ bin indicates luminosities excluded by the 
bright end magnitude limit of the survey. Each panel at $z > 0.2$ shows a Schechter 
function fitted to the ``minimal'' luminosity function (solid curve),
the fit for the luminosity function at $0.2 < z < 0.5$ (dashed curve)
and the Loveday et al. 
(1992) overall luminosity function, which is not split by color (dotted curve).
(b) As in (a) except that the ``best estimate'' luminosity function (see text) is 
shown.
(c) As in (a) and (b) except that the ``best estimate'' luminosity function is 
calculated assuming $q_0$ = 0.0}

\end{figure}

\begin{figure}

\caption{Left: the CFRS ``best estimate'' luminosity
function split by color for $0.05 < z < 0.2$
and $0.2 < z < 0.5$. Solid lines are Schechter function segments fit to the 
blue and red data separately, and the sum of these. The dotted curve is the Loveday 
et al. (1992) luminosity function, which is not split by color.  Right: The color-dependent
luminosity function from Metcalfe et al. (1991) split at (B-V) = 0.72,
again compared with the
Loveday et al. (1992) luminosity function. There is good agreement between the two
color-dependent luminosity functions, which both show an excess over the Loveday et al.
(1992) faint end (see also Marzke et al. 1994a)}

\end{figure}

\begin{figure}

\caption{Error ellipses for $M_{AB}*(B)$ and $\phi *$ obtained by fitting a Schechter function with 
$\alpha = -0.5$ to the $q_0$ = 0.5 luminosity functions of red galaxies (the left hand panels of Figure 
3(b)) in the three redshift intervals $0.2 < z < 0.5$ (solid curve), $0.5 < z < 0.75$ (dotted 
curve) and $0.75 < z < 1.0$ (dashed curve). These show that the luminosity functions are 
consistent with either no change in either parameter or with a combination of modest 
brightening and decreasing comoving density with increasing redshift. The vectors emanating
from each ellipse show the effect of changing $q_0$ to $q_0 = 0$}

\end{figure}

\begin{figure}

\caption{Rest-frame color-magnitude diagrams for the CFRS sample. Objects with open 
symbols have estimated redshifts (see text). At constant luminosity (e.g. above the 
horizontal line) the increase in luminosity of the blue galaxies relative to the red galaxies 
at $z > 0.5$ is apparent.}

\end{figure}

\begin{figure}

\caption{The upper panel shows the overall luminosity function for all CFRS galaxies in the $0.05 
< z < 1.0$ redshift range. Circles with error bars are the ``minimal'' luminosity function and 
the crosses represent the ``weighted'' luminosity function (see text). The solid and dashed 
curves represent fitted Schechter functions (best fit and 1$\sigma$ fits) and the dotted curve 
represents the Schechter function found for local galaxies by Loveday et al. (1992). The 
four lower panels show the ``best estimate'' luminosity function (see text) split in to four 
redshift bins. Redshift range and number of objects in that range are indicated by the 
label in the upper left of each panel. Galaxies to the left of the vertical line in the $0.05 < z 
< 0.2$ bin were excluded by the bright end magnitude limit of the survey. In each panel at 
$z > 0.2$, a Schechter function fit to the luminosity function is shown 
(solid curve) along with
the fit for the luminosity function at $0.2 < z < 0.5$
(dashed curve)
and the Loveday et al. (1992) overall luminosity function (dotted curve).}

\end{figure}

\begin{figure}

\caption{Variation in $M_{AB}(B)$ with redshift in the Steidel et al. (1995) sample of absorption 
line selected galaxies, divided into redder galaxies (open symbols) and bluer galaxies 
(filled symbols). The bluest galaxies are represented by filled squares. The lines indicate 
linear regressions of $M_{AB}(B)$ on $z$ for these three samples (solid line - red, dotted - bluer, 
dashed - bluest). Although of marginal significance in this small sample, 
the constancy in the luminosity of 
red galaxies and the increase in the luminosity of bluer galaxies is as found in the present 
work.}

\end{figure}

\begin{figure}
  
\caption{Number magnitude counts in the $I$-band and $B$-band obtained by replicating the 
observed CFRS population to higher redshifts
(connected open symbols), 
compared with observations (taken from LCG - solid symbols). There is a 
modest shortfall in the $I$-band but a substantial one in the $B$-band indicating further 
evolution in the luminosity function from blue galaxies will be required.}

\end{figure}

\begin{figure}

\caption{The $N(z)$ distributions obtained from the CFRS luminosity function in the $17.5 < 
I_{AB} < 22.5$ range (upper histogram) and extrapolated, assuming no further evolution, to 
fainter magnitudes, $22.5 < I_{AB} < 23.5$ and $23.5 < I_{AB} < 24.5$.}

\end{figure}

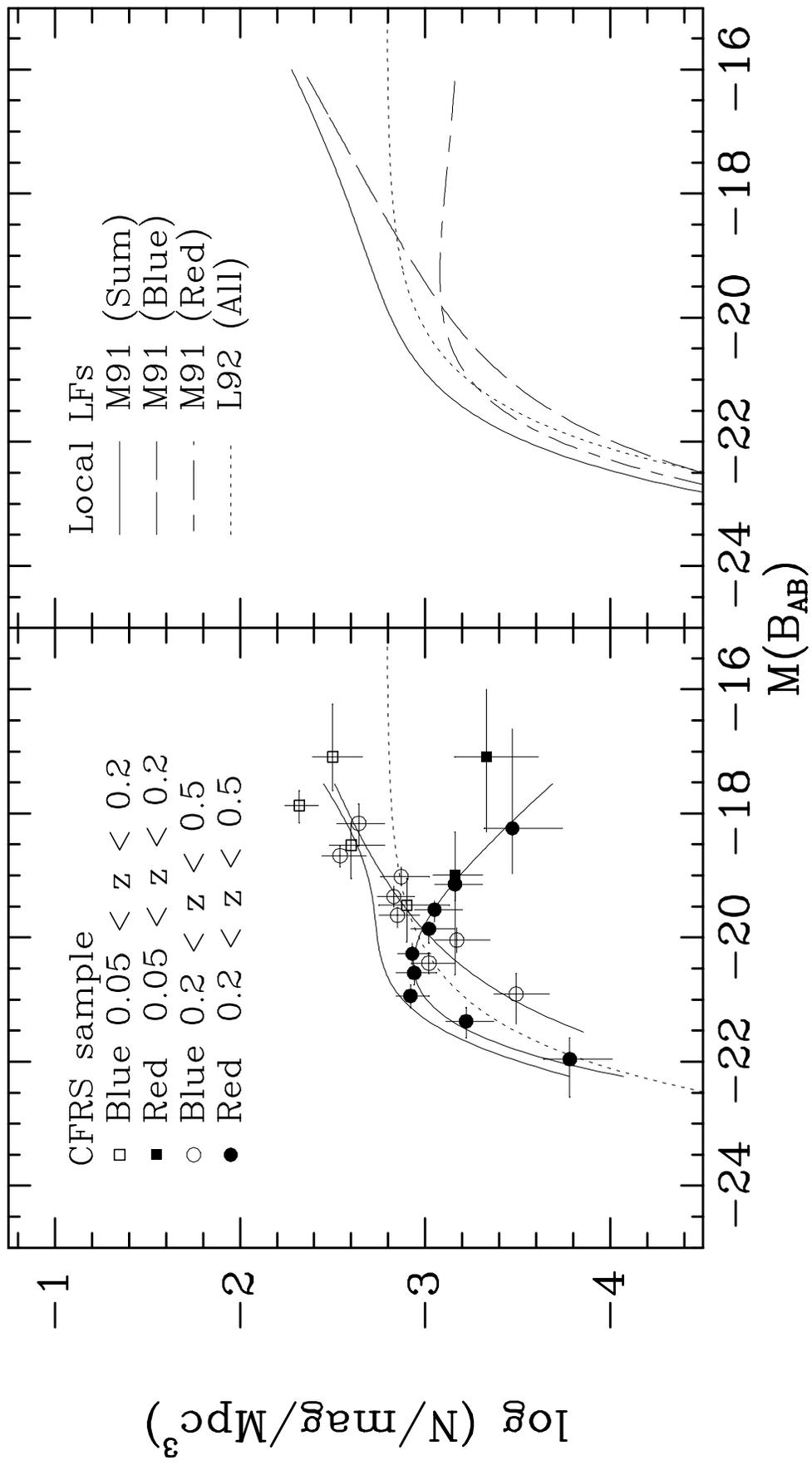
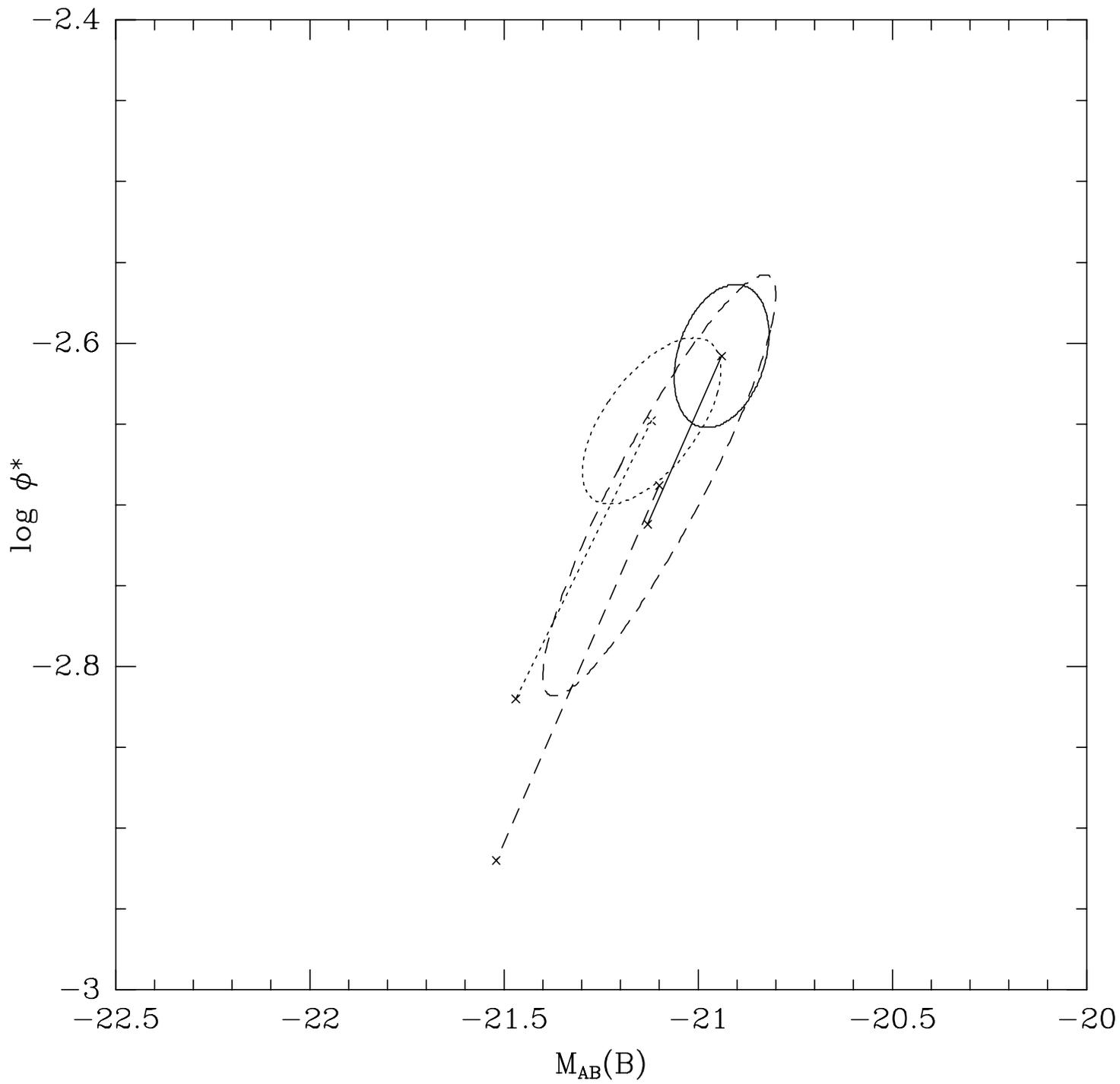
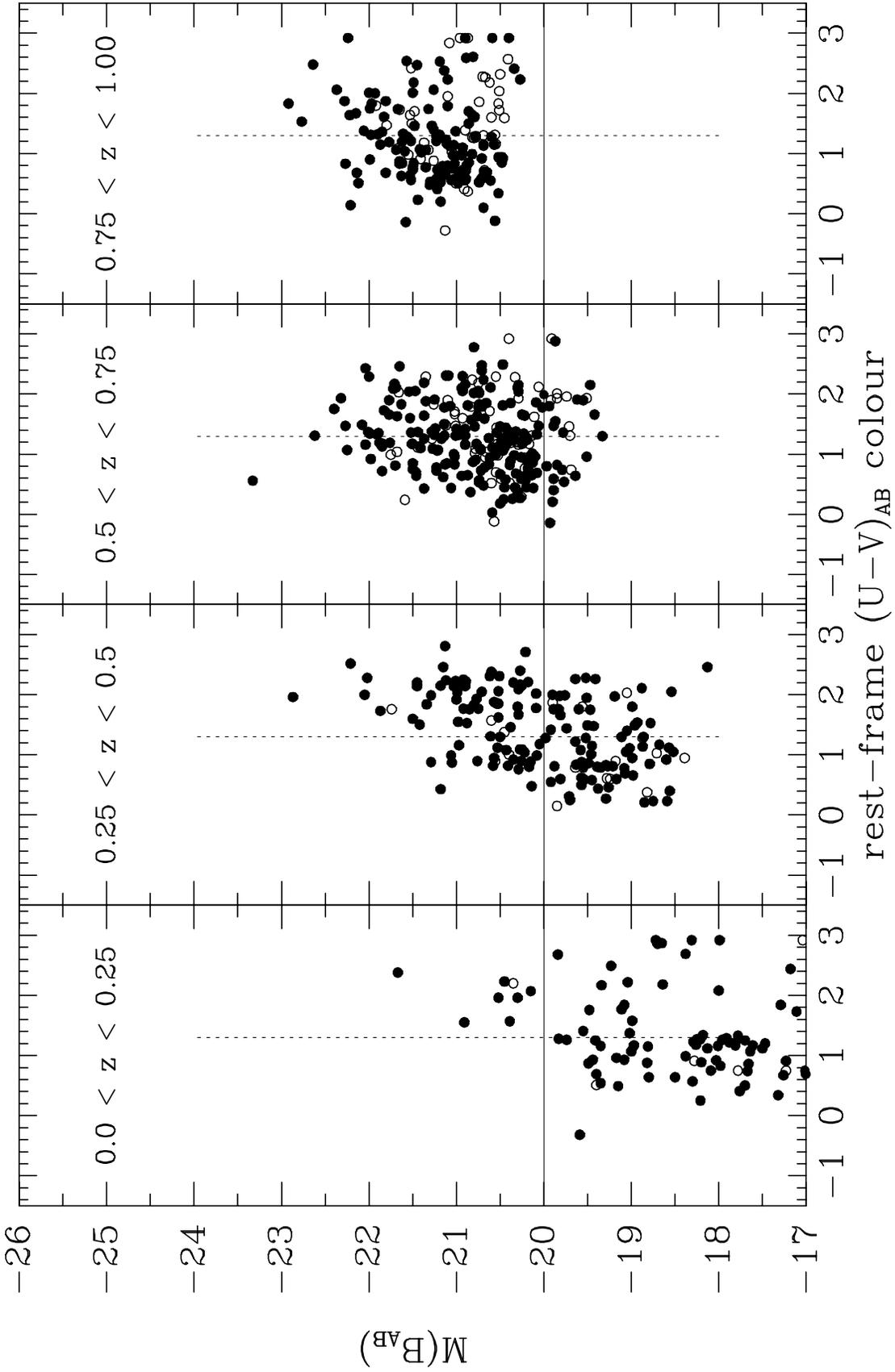
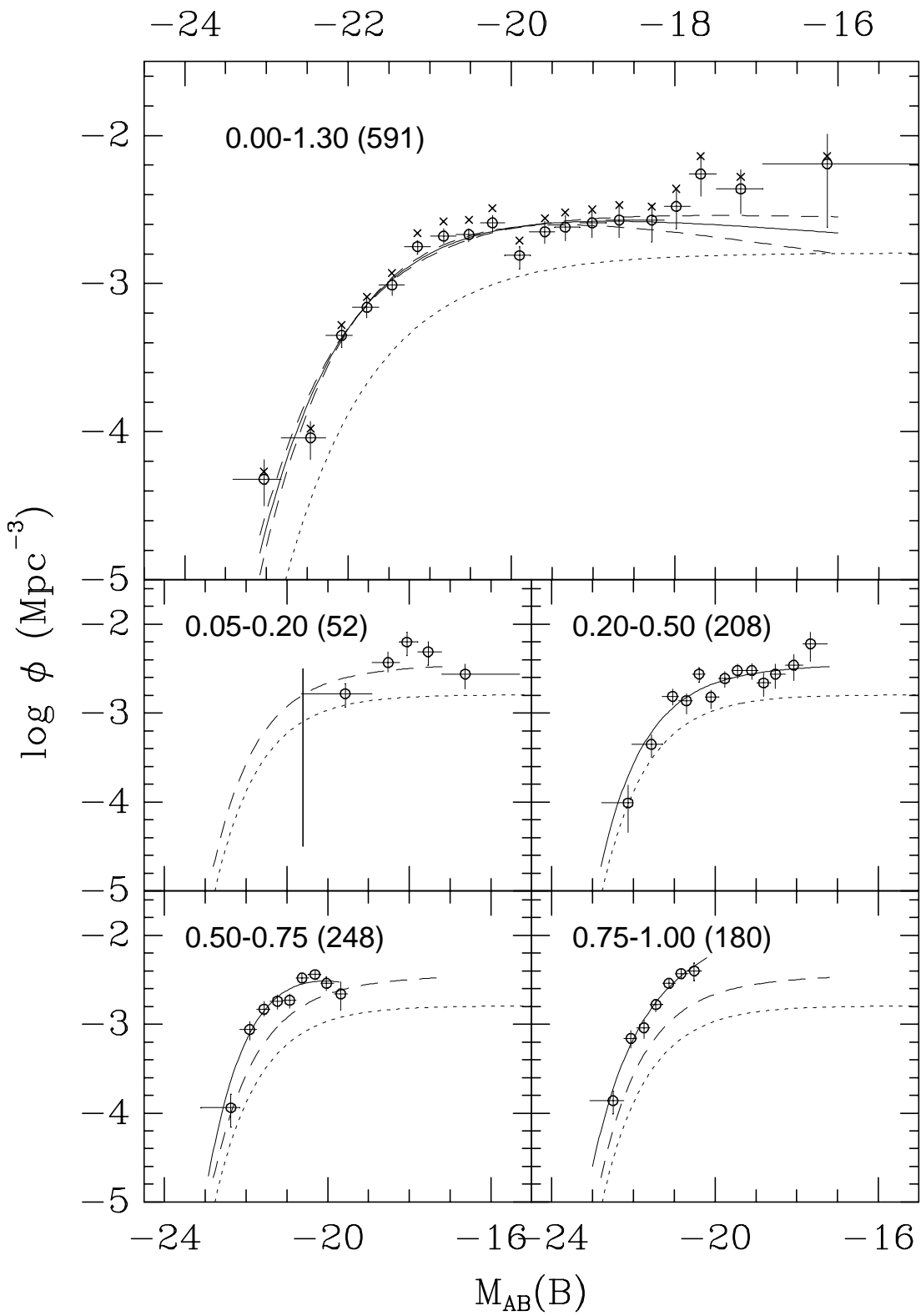
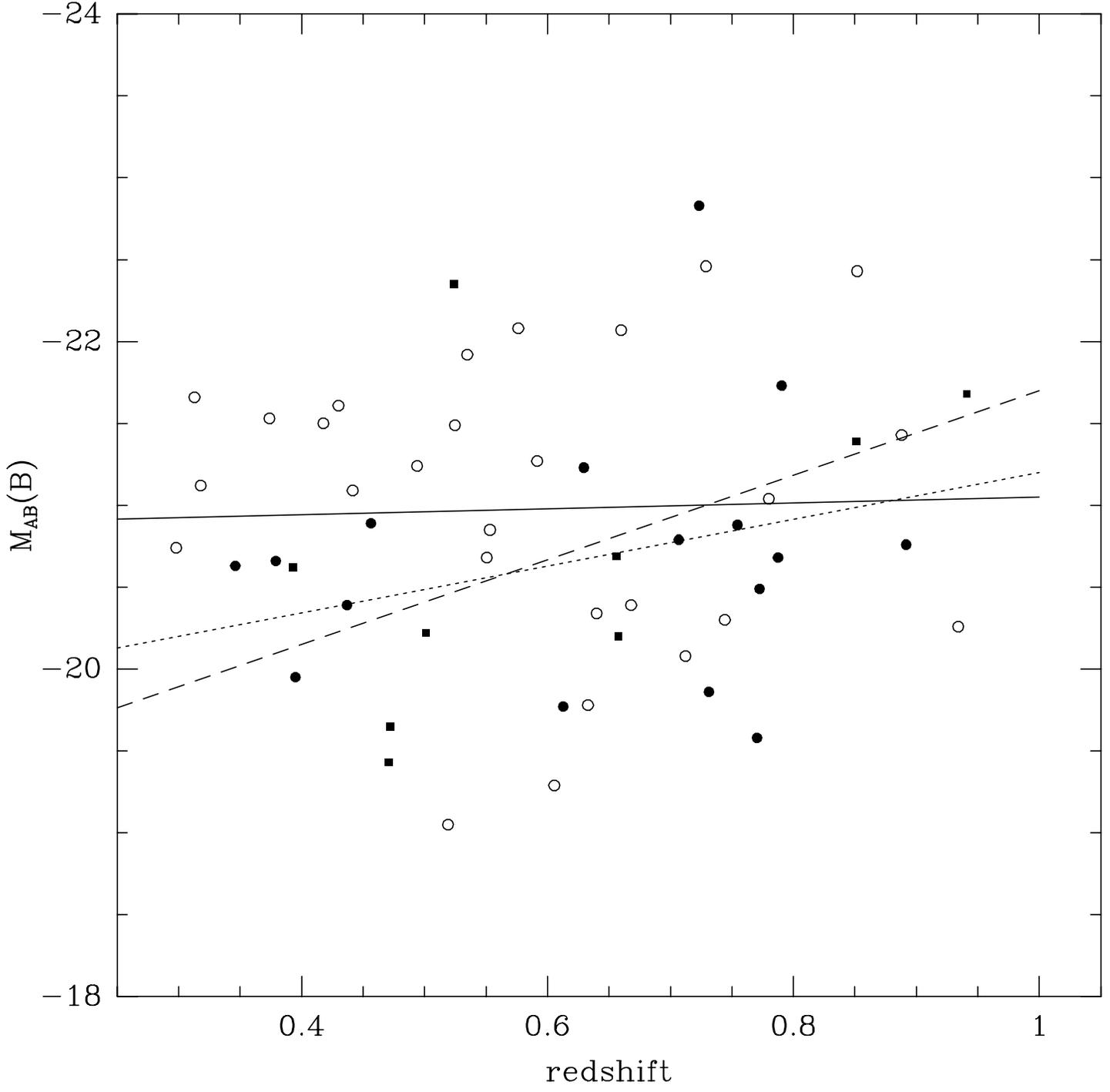
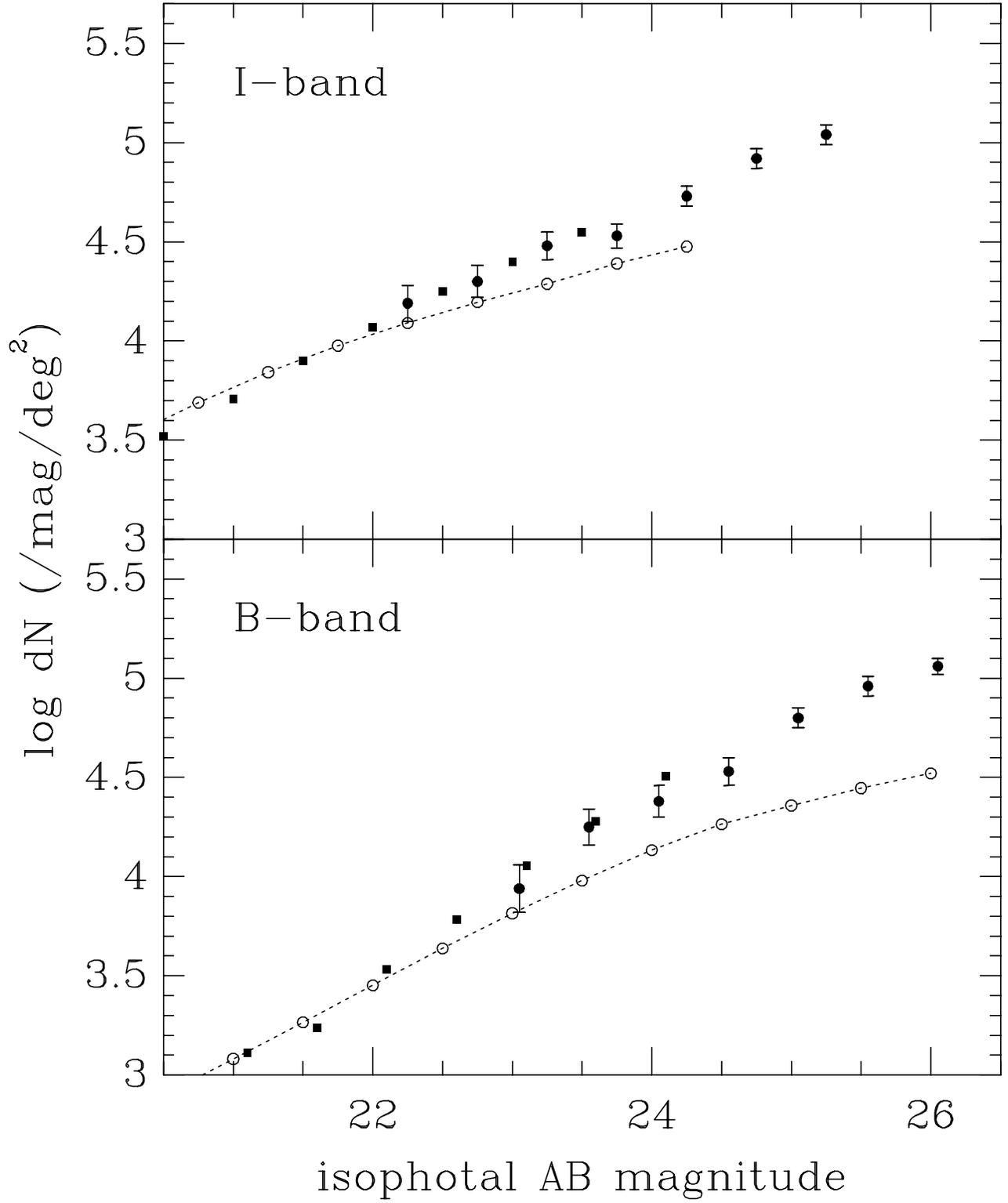
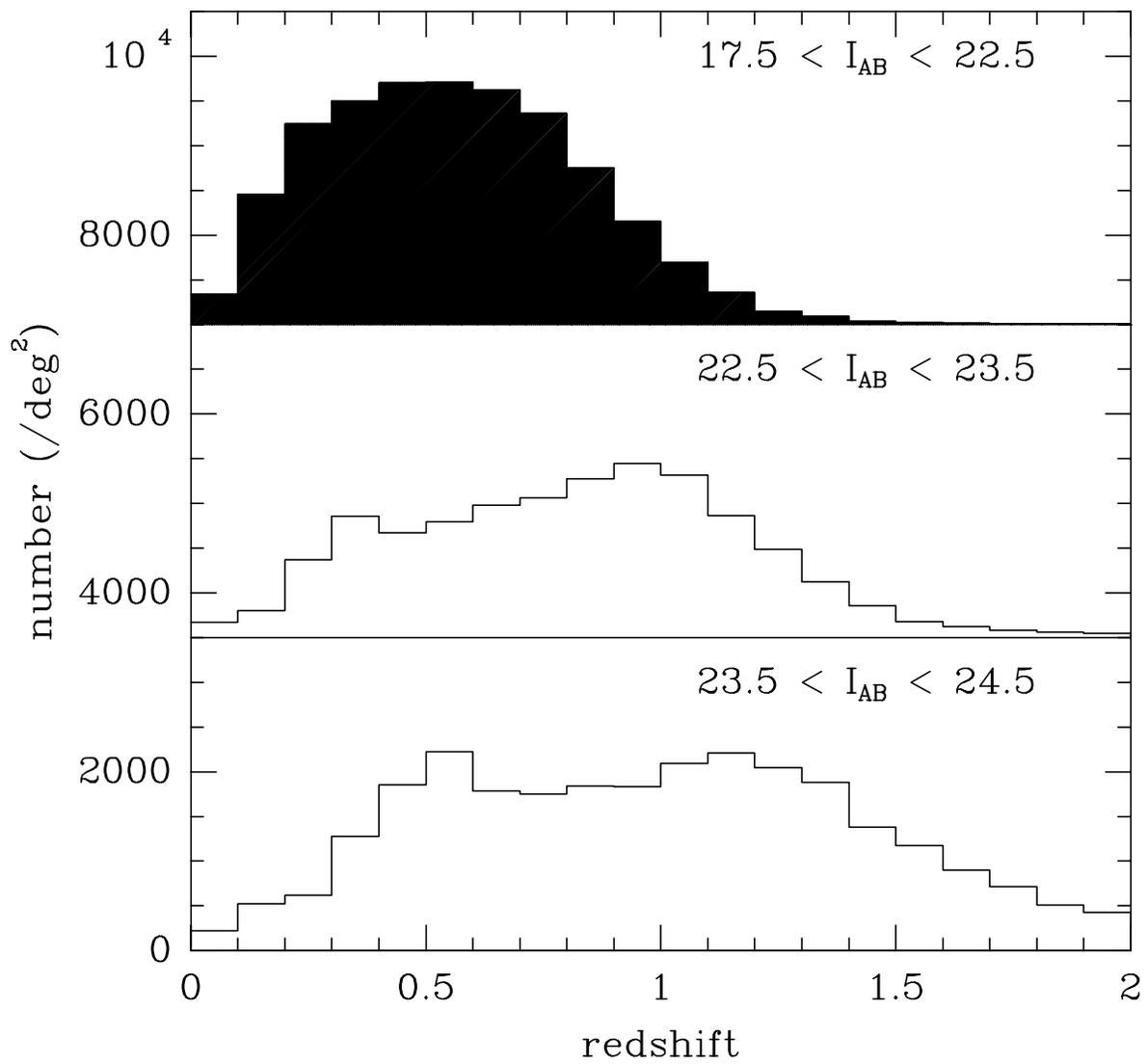
\end{document}